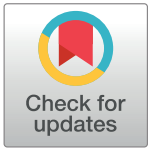

RESEARCH ARTICLE

# Optimal occlusion uniformly partitions red blood cells fluxes within a microvascular network

**Shyr-Shea Chang[1]\*, Shenyinying Tu[1], Kyung In Baek[2], Andrew Pietersen[2], Yu-Hsiu Liu[3], Van M. Savage[4,5,6], Sheng-Ping L. Hwang[7], Tzung K. Hsiai[2,8], Marcus Roper[1,4]**

**1** Department of Mathematics, University of California Los Angeles, Los Angeles, California, United States of America, **2** Department of Bioengineering, School of Engineering & Applied Science, University of California Los Angeles, Los Angeles, California, United States of America, **3** Department of Life Science, National Taiwan University, Taipei, Taiwan, Republic of China, **4** Department of Biomathematics, University of California Los Angeles, Los Angeles, California, United States of America, **5** Department of Ecology and Evolutionary Biology, University of California Los Angeles, Los Angeles, California, United States of America, **6** Santa Fe Institute, Santa Fe, New Mexico, United States of America, **7** Institute of Cellular and Organismic Biology, Academia Sinica, Nankang, Taipei, Taiwan, Republic of China, **8** Division of Cardiology, Department of Medicine, University of California Los Angeles, Los Angeles, California, United States of America

\* sdawn0728@gmail.com

## Abstract

In animals, gas exchange between blood and tissues occurs in narrow vessels, whose diameter is comparable to that of a red blood cell. Red blood cells must deform to squeeze through these narrow vessels, transiently blocking or occluding the vessels they pass through. Although the dynamics of vessel occlusion have been studied extensively, it remains an open question why microvessels need to be so narrow. We study occlusive dynamics within a model microvascular network: the embryonic zebrafish trunk. We show that pressure feedbacks created when red blood cells enter the finest vessels of the trunk act together to uniformly partition red blood cells through the microvasculature. Using mathematical models as well as direct observation, we show that these occlusive feedbacks are tuned throughout the trunk network to prevent the vessels closest to the heart from short-circuiting the network. Thus occlusion is linked with another open question of microvascular function: how are red blood cells delivered at the same rate to each micro-vessel? Our analysis shows that tuning of occlusive feedbacks increase the total dissipation within the network by a factor of 11, showing that uniformity of flows rather than minimization of transport costs may be prioritized by the microvascular network.

## Author summary

Arterial trees shuttle red blood cells from the heart to billions of capillaries distributed throughout the body. These trees have long been thought to be organized to minimize transport costs. Yet red blood cells are tightly squeezed within the finest vessels, meaning that these vessels account for as much as half of the total transport costs within the arterial network. It is unclear why vessel diameters and red blood cell diameters are so closely

Institutes of Health (NIH), under a Ruth L. Kirschstein National Research Service Award (T32-GM008185, http://www.biomath.ucla.edu/huang/sibtp/). TKH received funding from National Institute of Health (NIH) under grant numbers R01 HL083015, R01 HL129727, and R01 HL111437 (https://www.nih.gov/). SPLH received funding from MOST (Ministry of Science and Technology, Taiwan, under grant MOST 105-2313-B-005-MY3, https://arspb.most.gov.tw/NSCWeb/wSite/mp?mp=11&token=ZVj37+cQGrJIZINi/). VMS received funding from the US National Science Foundation (ID: 1254159, http://nsf.gov/). The contents of this paper are solely the responsibility of the authors and do not necessarily represent the official views of the NIH. The funders had no role in study design, data collection and analysis, decision to publish, or preparation of the manuscript.

**Competing interests:** The authors have declared that no competing interests exist.

matched in a network that is presumed to optimize transport. Here, we use mathematical modeling and direct observations of red blood cell movements in embryonic zebrafish to show that occlusive feedbacks—the pressure feedbacks that alter the flows into a vessel when it is nearly blocked by a red blood cell—can optimally distribute red blood cells through microvessels. In addition to revealing an adaptive function for the matching of vessel and red blood cell diameters, this work shows that uniformity of red blood cell fluxes can be a unifying principle for understanding the elegant hydraulic organization of microvascular networks.

## Introduction

Vascular networks transport oxygen, carbon dioxide and sugars within animals. Exchange of both nutrients and gases occurs primarily in narrow vessels (e.g. capillaries) that are typically organized into reticulated networks. The narrowest vessels are comparable in diameter to red blood cells, forcing cells to squeeze through the vessels. Accordingly, hereditary disorders or diseases affecting the elasticity of cells and preventing them from contorting through narrow vessels can disrupt microvascular circulation [1]. The cost of blood flow transport in the cardiovascular system is thought to dominate the metabolic burden on animals [2]. The rate at which energy must be expended to maintain a constant flow of blood through a vessel is inversely proportional to the 4th power of the vessel radius. Red blood cells occlude the vessels that they pass through, further increasing the resistance of those vessels [3]. Accordingly capillaries and arterioles account for half of the total pressure drop within the network, and thus half of its total dissipation [4]. Experiments in which cells are deformed using optical tweezers, or by being pushed through synthetic micro-channels have shown that the extreme deformability of mammalian red blood cells requires continous ATP powered-remodeling of the connections between membrane and cytoskeleton. ATP released by deformed cells may induce vasodilation facilitating passage of cells through the narrowest vessels [5]. Thus, chemical as well as hydraulic power inputs are needed to maintain flows through microvessels [6, 7].

Why do micro-vessels need to be so narrow? A textbook answer to this question is that smaller, more numerous capillaries allow for more uniform vascularization of tissues—ensuring that "no cell is ever very far from a capillary" [4]. If smaller vessels are favored physiologically and red blood cell diameter acts as a lower bound on capillary diameters, then networks in which capillary diameters match those of red blood cells may be selected for. However, red blood cell sizes do not seem to be stiffly constrained—for example measured red blood cell volumes vary over almost an order of magnitude (19 to 160 femto-liters) between different mammals [8]. Since for a fixed capillary diameter, a small decrease in red blood cell diameter would greatly reduce rates of energy dissipation for red blood cells traveling through capillary beds [9], the evolutionary forces maintaining red blood cells and capillary diameters remain unclear.

There is a natural analogy between occlusion of vessels by red blood cells, and the congestion that occurs in data or road networks [10, 11]. Efforts to construct efficient transport networks often focus on reducing congestion [10], yet although cardiovascular networks are thought to be organized to minimize transport costs (i.e. the viscous dissipation occuring within the network) [12, 13]; the presence of congestion at the finest scales seems at odds with minimizing these costs. Could the extreme deformation of cells passing through capillaries be an adaptive feature of the cardiovascular network? By directly stretching cells using optical tweezers Rao et al. [14] showed that deforming red blood cells releases oxygen. But it remains an untested hypothesis that squeezing cells so that they may pass through capillaries accelerates oxygen release, and therefore contributes to the function of the network. Indeed, earlier models suggest that alterations in the shape of the red blood cell surface decrease rates of oxygen exchange [15].

In this work we use mathematical modeling to reveal a previously unreported contribution of occlusive dynamics to the efficient functioning of the cardiovascular network. Moreover we link occlusive dynamics to a different open mystery of cardiovascular function. Specifically given that microvessels are distributed throughout the body and at very different distances from the heart, there is surprising consistency among measured flow rates in different capillaries [16–18] (with some exceptions [19]). Indeed consistency in flow rates may be biophysically necessary: if flow rate in a capillary is too low, the cells surrounding the capillary may not receive enough oxygen, but if the flow rate is too high, then red blood cells may leave the capillary bed before surrendering their oxygen to the surrounding cells. If the cardiovascular system is treated as an idealized symmetric branching network (such as in [2]) then flows are automatically uniformly partitioned at each level of the network, including among capillaries. But real cardiovascular networks have complex topologies, and it is not clear how the uniform flow can be achieved among billions of capillaries whose distances from the heart can range over several orders of magnitude.

In this work we show that in the embryonic zebrafish, a model system for studying cardiovascular development [20], answers to these two questions may be closed linked. Tuned occlusion—i.e. small differences in the resistance that vessels present to cells—ensures that red blood cells are uniformly partitioned between the finest vessels within the zebrafish trunk. Although zebrafish red blood cells have quite different morphologies from mammalian red blood cells, the matching in sizes of red blood cell and narrow vessel means that occlusive dynamics occur in the zebrafish network. Our experimental observations confirm previous measurements that red blood cells are uniformly partitioned between fine vessels [18], yet in the absence of tuned occlusion, we demonstrate that the vessels closest to the zebrafish heart would receive 11-fold higher rates of flow that vessels furthest from the heart. In other words these vessels would act as hydraulic short-circuits. In further support of the hypothesis that occlusion is an adaptive feature of the network we calculate optimal occlusive dynamics—i.e. the distribution of occlusive feedbacks (the negative feedbacks each cell exerts on cells trying to enter the same vessel) that leads to the most uniform partitioning of red blood cells between the smallest vessels. The occlusive feedbacks within the real zebrafish conform very closely to this optimal distribution.

Microvascular networks have been postulated to be organized to minimize the cost of transport (i.e. the total viscous dissipation associated with blood flow) [12, 21–23]. Certainly in larger vessels within both the arterial and venous vascular network, vessel radii appear to be organized to minimize dissipation [13, 24]. Yet, our results suggest that rather than eliminating cellular congestion, fine vessels make use of it. As a direct demonstration of the tradeoff between minimizing the cost of transport and tuning occlusion to route red blood cells uniformly, we show that the optimal distribution of occlusive feedbacks that uniformizes red blood cell partitioning increases hydraulic dissipation in the network 11 fold compared with a network in which the smallest measured occlusive feedbacks occur within each vessel. Thus, taken together, our results advance a potential new optimization principle—uniform routing of red blood cells—that may underlie the organization of microvascular networks generally.

## Materials and methods

### Ethics statement

All animal experiments performed at Academia Sinica were approved by the Animal Use and Care Committee of Academia Sinica (protocol # 12-12-482). Zebrafish were bred and maintained at the UCLA Core Facility. Zebrafish experiments were performed in compliance with the Institutional Animal Care and Use Committees (IACUC) at the University of California, Los Angeles (UCLA) (under animal welfare assurance number A3196-01)

### Imaging zebrafish trunk vessels and red blood cell movements

To measure the red blood cell fluxes in zebrafish trunk vascular network we cultured double transgenic Tg(fli1:GFP; gata1:ds-red) zebrafish embryos, in standard E3 medium supplemented with 0.05% methylene blue solution at 28.5˚C. In this transgenic fish line, fli1, a transcription factor associated with blood vessel morphogenesis is tagged with green fluorescent protein, causing the endothelial cells surrounding blood vessels to fluoresce green. Additionally, GATA-1, a transcription factor associated with erythrogenesis is tagged with red fluorescent protein, so that the red blood cells traveling through the GFP-labelled network fluoresce red. Zebrafish larvae were sedated with neutralized 0.02% tricaine solution(Sigma, MO) and mounted in 1-2% low melting agarose (Sigma-Aldrich, MO) for imaging. Erythrocytes were imaged at 4 day post fertilization (dpf) under a fluorescent microscope (Zeiss, Germany) with 50 ms exposure time. To measure detailed geometry and occlusive feedback of zebrafish trunk network we re-imaged a single 4 dpf zebrafish. We measured vessel lengths and radii from GFP-images taken under 10× magnification using a Zyla sCMOS camera on a Zeiss Axio Imager A2 fluorescent microscope. To measure the flow velocity, the same scope was used to take images in the DsRed channel at time intervals of 0.078 – 0.107 sec. Red blood cells were manually tracked in image sequences using ImageJ [25].

### Mathematical modeling of occlusion and parameter estimation

Flow is laminar within each zebrafish microvessel [26, 27]. The Womersley number [28] that characterizes the importance of unsteadiness effects in time-dependent flow, which for a vessel of diameter $d$, carrying blood with kinematic viscosity $v$, and with heart rate $f$, is given by $Wo = \sqrt{\frac{2\pi f d^2}{v}}$. Within the largest trunk vessels $d \approx 12\ \mu$m, the viscosity of whole blood is $v \approx 5 \times 10^{-6}$ m$^2$/s [29], and the heart-rate is approximately $f = 2$ s$^{-1}$, so $Wo = 1.9 \times 10^{-2} \ll 1$, meaning that we may neglect pulsatile effects. Flow is uniform along each vessel, except within an entry region whose length is $\ell \ll Ud^2/v$ for a vessel of diameter $d$, through which blood travels at a speed $U$ [30]. Maximum blood velocities are on the order of 0.3 cm/s [31], so using the diameter of the largest trunk vessels we obtain: $\ell \ll 0.3\ \mu$m. Since the entry region is much smaller than the typical vessel length, we treat the flow in each vessel as being uniform along its length. Putting these ingredients together, we find that the flow through each vessel is inversely proportional to the resistance of the vessel, and the resistance may be calculated using Stokes' equations (i.e. the equations for slow-creeping flows [30]) from the geometry of the vessel and from the number of red blood cells that it contains. Mechanistic models to predict the motions of red blood cells through micro-vessels or through microfluidic channels with comparable diameters have been developed in previous works [3, 32, 33]. Throughout this work we adopt a simple model for red blood cell occlusion in which the resistance of each vessel increases linearly with the number of red blood cells present. That is, if the number of red blood cells in a narrow vessel is given by $n$, then its resistance is given by an equation:

$$R(n) = R_0 + n\alpha_c .\qquad(1)$$

where $R_0$ is the resistance of the vessel in the absence of red blood cells, i.e. is given by the Hagen-Poiseuille law relating the pressure drop and flow rate in a tube carrying viscous fluid, so that for a vessel of length $\ell$ and radius $r$: $R_0 = \frac{8\mu_{pl}\ell}{\pi r^4}$, where $\mu_{pl} \approx 1 cP$ is the viscosity of the non-red blood cell component of the flood. Here the parameter $\alpha_c$, which we call the occlusion strength in this paper, gives the increase in vessel resistance per red blood cell. Eq (1) represents a form of non-Newtonian rheology, the deviation of resistance from simple viscous fluid. In particular, the apparent viscosity of blood, i.e. $R(n)\pi r^4/8\mu_{pl}\ell$, increases with hematocrit, i.e. with the concentration of red blood cells. Eq (1) can be derived from the micromechanical model of [34]. Indeed any model in which the pressure drop across the red blood cell is proportional to the velocity of the cell will produce a relationship like Eq (1), and so identical equations are also used to model the traffic of droplets or particles through microfluidic channels [35, 36]. In all of these models, $\alpha_c$, which we may think of as the intrinsic resistance of a single cell [34, 35, 37, 38], depends on the specific details of how the movements of cells, droplets or particles along the walls of the capillary or channel are lubricated. $\alpha_c$ therefore depends on parameters that we can not measure experimentally, including the thickness and porosity of the glyocalyx that coats the endothelial wall of the capillary, as well as being sensitive to changes in vessel radius [33, 34] that are too small to be detected in light microscopy. It also depends upon the elastohydrodynamic deformation of both the cells and the capillary wall [32]. Accordingly we treat $\alpha_c$ as a phenomenological constant, to be measured directly by fitting Eq (1) to real flow data. Specifically for each micro-vessel, we can measure both the velocity of flow within the vessel and the number of red blood cells that it contains. We note that due to the Fahraeus effect [36, 39] the velocity of red blood cells is in general larger than the flow velocity. However in human vessels whose diameters are comparable relative to human red blood cells to the diameter of the zebrafish vessel relative to the zebrafish's red blood cells, the ratio of red blood cell velocity to whole blood velocity is less than 1.09 [39]. Hence we approximate the flow velocity by the velocity of the red blood cell in this measurement. The pressure difference across each vessel varies in time due to the variable pressure within the aorta, and also, less predictably because, since the resistance of all vessels changes from moment to moment, there are pressure feedbacks across the entire network. But we assume that there is an overall average pressure drop across each vessel that is constant in time but changes from vessel to vessel. Under conditions of time-independent pressure drop, the velocity of cell movement, $v$, in each vessel will be inversely proportional to the vessel resistance $R(n)$. Thus Eq (1) predicts that a plot of $1/v$ against $n$ will give a straight line, the slope of which can be used to calculate $\alpha_c$. Here we used the modeled flows in the fine vessels where no red blood cell is present to determine the intercepts, which can be calculated by using Hagen-Poiseuille formula (see Results, Absence of occlusion . . .). By regressing $1/v$ against $n$ for each micro-vessel we calculate the variation of occlusive effects through the network (see S1 Text for more details of the regression).

### Incorporating occlusion into transport models

To study how varying occlusive effects between different microvessels may affect distribution of red blood cells, we incorporated Eq (1) into both continuum and discrete models of transport through the network.

For continuum level modeling, we assumed that the concentration of red blood cells was a constant, $\rho$, in each vessel. *Phase separation* of red blood cells can occur when flows divide at vessel junctions—that is red blood cells may split in different proportions than whole blood [40]—but separation was not seen in our data (i.e. all Se vessels had the same average red blood cell concentration of number per volume), and cannot account for the uniformity of red blood cell flows, as we discuss in the Results section. Thus if the constant concentration (number/volume) of red blood cells is $\rho$, then a vessel of volume $V$ is expected to contain $n = \rho V$ cells. Once each vessel in the network has been assigned a resistance, then we can solve for the flows in the entire network, by applying Kirchoff's first law (conservation of flux) to calculate the pressure at each branching and fusion point, and then using the pressure difference across each vessel to calculate flows [12, 41, 42]. We discuss the geometry of the network and boundary conditions in the Results section.

Since each micro-vessel is so small, typically each vessel contains no more than one or two cells at a time (but occasionally 3-5 cells were present in a vessel, see S2 Fig). For this reason we expected Poisson noise effects (i.e. fluctuations in the number of cells contained within each vessel) to influence red blood cell fluxes. We therefore built a discrete model, in which the trajectories of every single red blood cell traveling through the trunk network were directly simulated. Our discrete model is based on the droplet traffic model of [35]. Initially 990 cells are distributed uniformly through aorta according to measured zebrafish red blood cell concentrations [43]. At each step we calculate the resistance for each capillary by Eq (1), and then use the hydraulic resistor network model to calculate the whole blood flow rates within each vessel. We then let cells travel according to the predicted whole blood velocity in their vessel. Again we assume that the velocity of cell matches with flow velocity in Se vessels. The diameter of the dorsal aorta (DA) is larger and this mismatch may be significant in the DA. Since the cell velocity depends linearly on the flow velocity we expect this effect to increase the cell fluxes in all Se vessels equally and to therefore influence the partitioning of cells only weakly. While for precise prediction of cell fluxes the inclusion of this velocity mismatch will be necessary, here we are developing a minimal model that singles out the effect of occlusive feedbacks, and hence we assume that the cell velocity is the same as flow velocity in all vessels. When a cell arrives at a node of the network; i.e. at a point where a vessel branches into two, which vessel it enters is determined randomly by a Bernoulli process; that is the probability of cell entering a vessel is determined by the flow rate ratio of the two vessels. We therefore suppress the Zweifach-Fung effect [44]. The Zweifach-Fung effect characterizes the uneven distribution of red blood cells at a branching point, depending, amongst other factors, on stream lines at the branching point, and exibility of the cell [45–47]. Here we use a minimal model that neglects the Zweifach-Fung effect because we see that only occlusive feedbacks can account for uniform partitioning of cells. Indeed, we found no difference between the red blood cell concentration concentration (number / unit volume) of vessels in the rostral Se artery ($2.88 \times 10^{-4} \pm 2.19 \times 10^{-4}$ $1/\mu m^3$) and in the caudal Se artery ($2.18 \times 10^{-4} \pm 2.72 \times 10^{-4}$ $1/\mu m^3$). Flows are then recomputed for the new distribution of cells. Cells that leave the network, i.e. reach the end of one of the vessels within the simulated part of the network are immediately reintroduced into the network via the aorta. For each combination of parameters, we simulated 1000 s of red blood cell movement, with a time step of 0.1 s. Using fluorescence microscopy to track red blood cells meant that our measurement frame rate was too low to directly measure cell velocities within the aorta. So we fit total inflow into the trunk via the aorta to match the mean flux across all fine vessels to the experimentally measured mean flux.

## Results

### Geometry of the zebrafish trunk microvasculature

The 4 day post fertilization zebrafish trunk vasculature is topologically simple. Oxygenated red blood cells (henceforth RBCs) flow into the zebrafish trunk via the dorsal aorta (DA) and return the heart via the posterior cardinal vein (PCV). The microvasculature consists of a series of parallel intersegmental vessels (Se) that, if the vasculature were laid flat, would span between the aorta and cardinal vein like the rungs of a ladder (Fig 1A). Se are divided into intersegmental arteries (SeA) that connect to the aorta, and intersegmental veins (SeV) that connect to the posterior cardinal vein. SeA and SeV connect via another vessel called the Dorsal Longitudinal

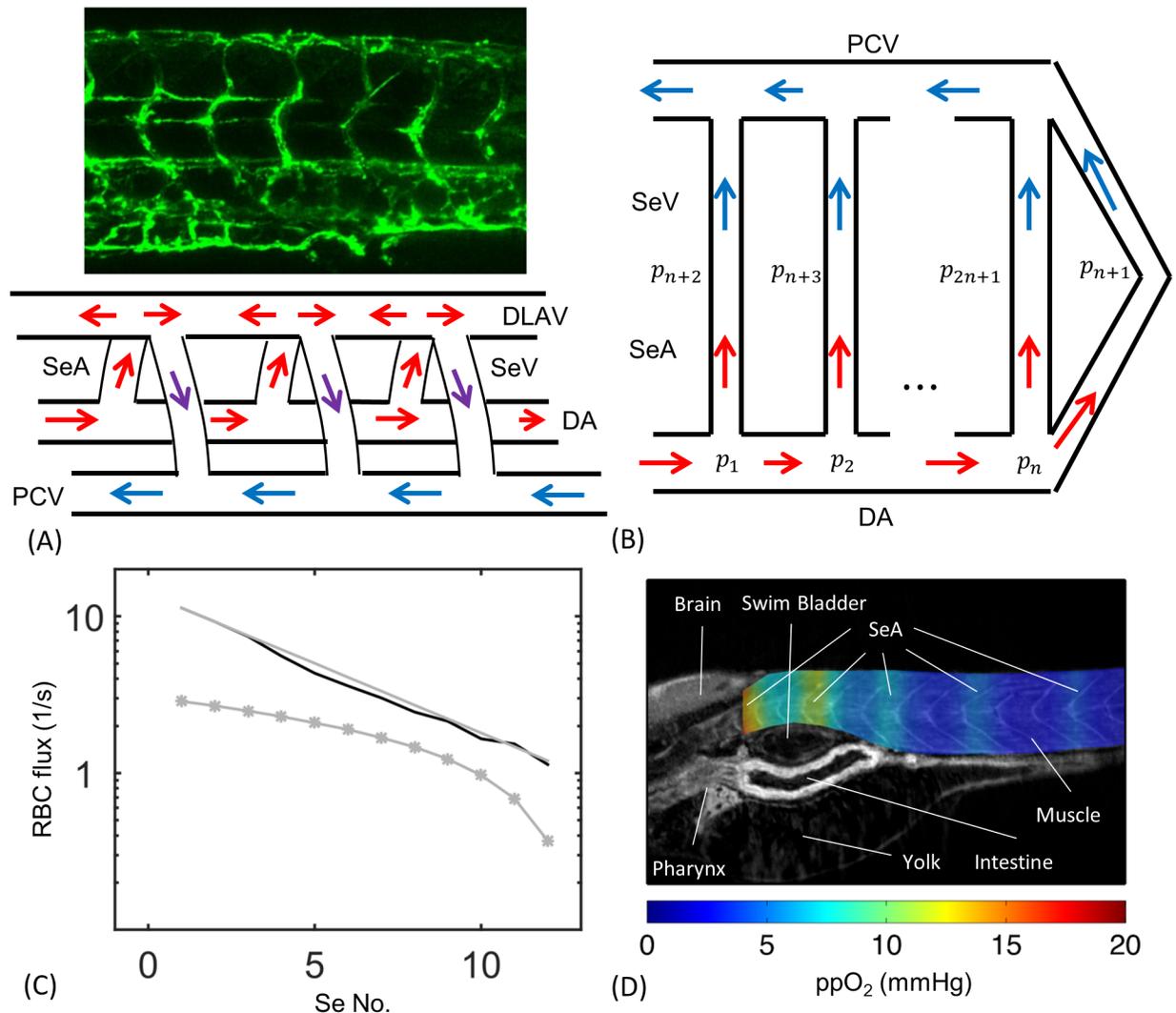

**Fig 1. The embryonic zebrafish trunk is perfused by a series of parallel intersegmental arteries (SeAs).** Hydraulic models for the network predict that the first of these SeA will short circuit flow through the trunk. (A) 4 day post-fertilization zebrafish embryo trunk network and wiring diagram showing PCV, DA and Se vessels in which SeA connect directly to SeV. (B) Representation of the same network as a set of hydraulic resistors. (C) A resistor network model predicts that cell fluxes decrease exponentially with distance from the heart (Black curve: numerical solution using real geometric parameters, Gray line: asymptotic model. For these two curves flow rates are multiplied by the concentration of red blood cell $\rho = 0.003$ $\mu m^{-3}$ measured in [43]). By contrast an occlusive feedback model incorporating uniform occlusion strength $\alpha_c = 1.01 \times 10^{-6}$ $g/\mu m^4$ $s$ did not lead to more uniform distribution of red blood cell fluxes between vessels (Gray stars). (D) Anisotropic fluxes produce uneven oxygen perfusion within the trunk. Simulation results are superimposed on a zebrafish CT image reproduced from [53].

https://doi.org/10.1371/journal.pcbi.1005892.g001

Anastomotic Vessel (DLAV), and in different parts of the DLAV, red blood cells flow toward the tail of the fish or toward its head. Red blood cells can enter the PCV by flowing along one of the SeAs, through a section of the DLAV, and then along one of the SeVs. Significantly, however, they can also flow directly from the DA into the PCV, since the two connect at the far end of both vessels in the tail of the fish.

The positions of SeAs and SeVs vary from embryo to embryo [48]. In particular, SeVs and SeAs do not strictly alternate their connections with the DLAV. To form a model that does not depend on any specific A-V pattern we choose to connect SeAs and SeVs directly in a pairwise manner (Fig 1B), reducing the model to a bilaterally symmetric network in which no flow occurs in the DLAV (which can therefore be suppressed). Then we assign the same conductances for directly connected SeAs and SeVs and the same conductances for sections of DA as for the symmetric matching segments of PCV. Under these symmetry assumptions the pressures at the intersection of SeA and SeV is the same for each SeA/SeV pair, and we can shift this pressure to zero without affecting the calculations. Solving flows in this network reduces to solving flows in the lower half of Fig 1B with fixed inflow in the beginning of the aorta and zero pressures at the intersections between SeAs and SeVs, and between DA and PCV at the tail.

### Absence of occlusion produces uneven fluxes within the SeA

As a first step we calculated the RBC flux in intersegmental arteries (SeA) with no occlusion or untuned occlusive effects and compared to experimental measurements. That is we approximated the resistance of each vessel using (1) with $\alpha_c = 0$ and treating the blood as a continuous phase, so that $\mu_{pl}$ replaced by $\mu_{wb}$, the viscosity of whole blood ($\mu_{wb} \approx 5$ cP in zebrafish [29]). This reduced model serves as a motivation and readers interested in the full model may skip to Results, Occlusive feedbacks with . . .. We measured the lengths of each vessel directly from fli1a-EGFP images. SeAs were all assigned the same radius (2.9 $\mu m$), while because the DA tapers from the head to the tail, we independently measured DA radii between each SeA (see S1 Table). Although ultimately tuned variation in SeA radii will be one way to explain changes in occlusive feedbacks, these variations strongly affect the parameter $\alpha_c$ in Eq (1) but have little effect on $R_0$. To model flows without feedbacks we can therefore neglect SeA radius variations. We focus on the arterial half of the network made up of SeA and DA vessels. We identify the vertices in this network, i.e. the points at which vessel branch or fuse, as points $i = 1, 2, \ldots n$, with respective pressures $p_i$ (Fig 1B). The number of SeAs, $n$, increases as the fish grows: for the 4 dpf zebrafish in our experiments $n$ ranges from 9 to 13. For definiteness in modeling, we assume $n = 12$. If vertices $i$ and $j$ are connected by a vessel, with resistance $R_{ij}$, then the total flow of blood along this vessel will be $(p_i - p_j)/R_{ij}$. Solving for the flows in the network is equivalent to finding the pressures $\{p_i\}$. For the zebrafish cardiovascular network we labeled vertices along the DA as $i = 1, 2, \ldots, n$. A vertex, $i = n + 1$, represents the end of the DA in the tail of the zebrafish, where it connects directly to the PCV, and we label the vertices where the SeA meet the DLAV as $i = n + 2, n + 3, \ldots 2n + 1$. At vertices $i = n + 1, \ldots 2n + 1$, our symmetry boundary conditions require that $p_i = $ const., and we set arbitrarily the value of this constant to be 0. Thus only the pressures $\{p_i : i = 1, \ldots n\}$ need to be determined. We find these pressures by applying Kirchoff's First Law (conservation of flux), at each point where the pressure is determined, i.e. $\sum_{j \in n(i)} (p_i - p_j)/R_{ij} = 0$, except at $i = 1$ (the vertex closest to the heart). At this vertex, $\sum_{j \in n(1)} (p_1 - p_j)/R_{1j} = F$, where $F$ is the total supply of blood to the trunk which is fit to real data (see Materials and methods). All summations are taken over the neighbor set, $n(i)$, i.e. over all vertices that are linked to $i$.

The model of the zebrafish trunk microvasculature as an hydraulic resistor network (neglecting occlusive effects) follows many previous capillary network models (see e.g. [12, 41, 42]). The equations are formally identical to those for an electrical resistor network, with pressures replacing voltages, and flow rates replacing currents. Just as placing a wire across the terminals of a battery in an electrical resistor network will short circuit the network (i.e. divert current from higher resistance paths), the first SeA is predicted to receive a larger-than-even share of the blood flow from the zebrafish trunk, with flow rates decreasing exponentially rapidly with distance from the heart. In total there is a predicted 11-fold difference between the flows through the first and last SeA (Fig 1C).

A simplified resistor network model that treats each SeA as having the same resistance, and assigns same resistances to each segment of DA between SeAs (i.e. ignores DA taper) quantitatively reproduces the exponential decay. To build the simplified model we assume that each segment of the DA has the same hydraulic resistance, and that each SeA has the same resistance. Using the measured mean radii and lengths, each DA has the same conductance, written as: $\kappa_1 = 1/R_1 = 9.4 \times 10^5$ $\mu m^4$ s/g, while all Se vessels have the same conductance, written as: $\kappa_2 = 1/R_2 = 3.9 \times 10^4$ $\mu m^4$ s/g. Then conservation of flow at vertex $i = 2, \ldots, n$ gives:

$$-\kappa_1 p_{i-1} + (2\kappa_1 + \kappa_2)p_i - \kappa_1 p_{i+1} = 0 ,  \qquad (2)$$

This is a second order recurrence equation with constant coefficients. Its general solution is:

$$p_i = C_+ \xi_+^i + C_- \xi_-^i, \qquad (3)$$

where $\xi_\pm$ are the roots of the auxiliary polynomial $\xi^2 - (2 + \lambda)\xi + 1 = 0$, in which there is a single dimensionless parameter: $\lambda = \frac{\kappa_2}{\kappa_1} = 0.04$. This equation has two roots, with $\xi_+ > 1$ and $\xi_- < 1$. In general $C_+$ and $C_-$ must both be non-zero to satisfy our boundary conditions (namely $p_{n+1} = 0$ and $F = \kappa_2 p_1 + \kappa_1(p_1 - p_2)$). However the two components give rise to exponentially growing and decaying pressures respectively. Typically the first term will negligible, except potentially in a small boundary layer region consisting of the vertices in the tail. Therefore over most vertices $p_i \sim C_- \xi_-^i$, i.e. the pressure decays exponentially with distance from the heart, causing flows in the SeAs to decay exponentially as a result. For the real zebrafish network: $\xi_- = 0.81$. Despite the simplification in geometry, the analytic formula agrees quite well with the solution to the full system of linear equations (compare gray and black curves in Fig 1C). Additionally, we note that for any $\lambda > 0$, it is impossible to organize an auxiliary polynomial without having one root $\xi_- < 1$, so exponential decay in fluxes is an inescapable feature of the ladder-like geometry of the trunk vasculature.

Although embryonic tissues receive oxygen primarily by diffusion through the skin [49, 50], vascular transport of oxygen becomes essential to embryo development after 2.5 weeks [51]. So we expect that a zebrafish with the large predicted difference in fluxes between trunk vessels would be disadvantaged. But because oxygen can diffuse through the zebrafish tissues, we first verified that the differences in fluxes predicted by the model lacking occlusive feedbacks would actually lead to differences in oxygenation in the trunk tissues. To do this, we modeled oxygen diffusion through the trunk by a reaction-diffusion equation, using the formulation and oxygen consumption coefficients derived by [52], and treating the vessels as oxygen sources (Fig 1D, and see S1 Text for details of the model). Note that our model includes only the contribution of oxygen perfusion from the blood to trunk oxygenation. For a real zebrafish at 4 dpf, these uneven oxygen levels would be compensated for by diffusion through the skin. However, our model shows that diffusion of oxygen within the zebrafish trunk can not compensate even at 4 dpf for uneven flows within the Se vessels.

### Red blood cell flows are uniform among trunk vessels

In contrast with the resistor network model, which predicts that the first Se vessel short circuits the network, measured RBC fluxes are nearly uniform between Se-vessels in living zebrafish. We tracked fluorescently tagged red blood cells moving through each of the 9 ∼ 13 SeAs within 6 living, sedated, zebrafish (see Materials and methods), over a total time interval of 26s per SeA. Fluxes in individual vessels varied greatly in time, due to the rapid change of blood pressures within the DA over the zebrafish cardiac cycle [31] and likely also due to nonlinear dynamics of the cells themselves within vessels [54], so the variability of flow rates was large for each vessel. However, mean fluxes varied little from vessel to vessel (Fig 2A). Each embryo

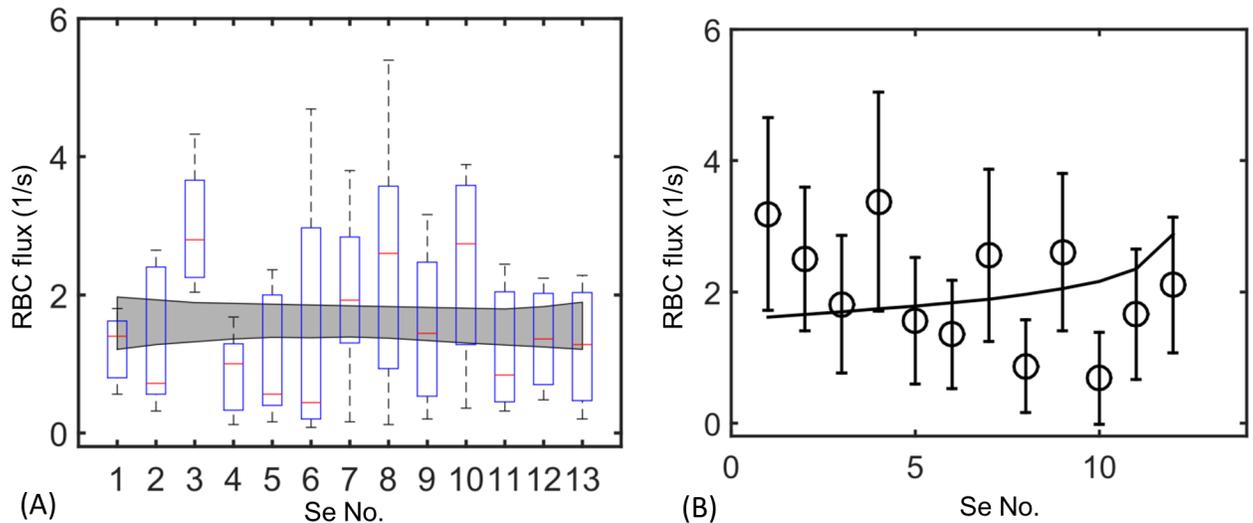

**Fig 2. Measured cell fluxes in real zebrafish embryos are almost uniform across all microvessels.** (A) Measured fluxes in 6 4dpf zebrafish. Box-and-whisker plots show the mean measured fluxes for all 6 zebrafish, while the gray region is the envelope produced by bootstrapped regressions of flux against Se No., which is a numbering of Se vessels starting from the rostral trunk. (B) A model incorporating tuned occlusion strength (black curve) agrees well with the data from a single 4dpf zebrafish (black circles), see Results, Tuning occlusive effects . . .. Bars: standard deviation on flux.

https://doi.org/10.1371/journal.pcbi.1005892.g002

exhibited variable RBC fluxes throughout the trunk. However the envelope of the lines of best fit for all six fish showed no consistent differences in RBC fluxes between first and last Se. Specifically from the six sets of zebrafish data we used bootstrapping method (generating replicate measurements for each Se vessel from the measured mean and standard deviation over all six fish) to estimate regression statistics. The gray envelope in Fig 2A shows the 95% confidence interval on all regressions thereby generated. We found that over all regressions $m = 0.012 \pm 0.032$ (mean ± standard deviation), showing no statistically changes in RBC flux from vessel to vessel.

## Occlusive feedbacks with variable strengths determine red blood cell fluxes

There are two major ingredients missing from the hydraulic resistor network model that could explain the anomalies between the predictions of that model and the real zebrafish flow rate data: phase separation of red blood cells and occlusive feedbacks effects [40, 55]. Separation occurs because red blood cells do not divide in the same ratios as whole blood when blood vessels branch: When a red blood cell passes through a junction at which a vessel branches into two daughter vessels of different sizes, it is more likely to enter the larger daughter vessel than would be expected based on the ratio of fluxes in the two daughter vessels. Phase separation cannot explain the uniform distribution of red blood cells seen across real zebrafish microvessels: to correct for an 11-fold difference in flow rates between first and last Se vessels, there would need to be an 11-fold increase in hematocrit between these vessels, in the absence of occlusive effects (since then hematocrit must increase exponentially to compensate for exponentially decreasing flow rates). This was not observed in our experiments. Indeed Pries et al. [33] explicitly fit measurements of red blood cell fluxes at the branch points of blood vessels, and parameterized the amount of phase separation that occurred. When we applied their model to the zebrafish microvasculature, only minute variations in hematocrit were predicted between different SeAs (see S1 Text and S1 Fig).

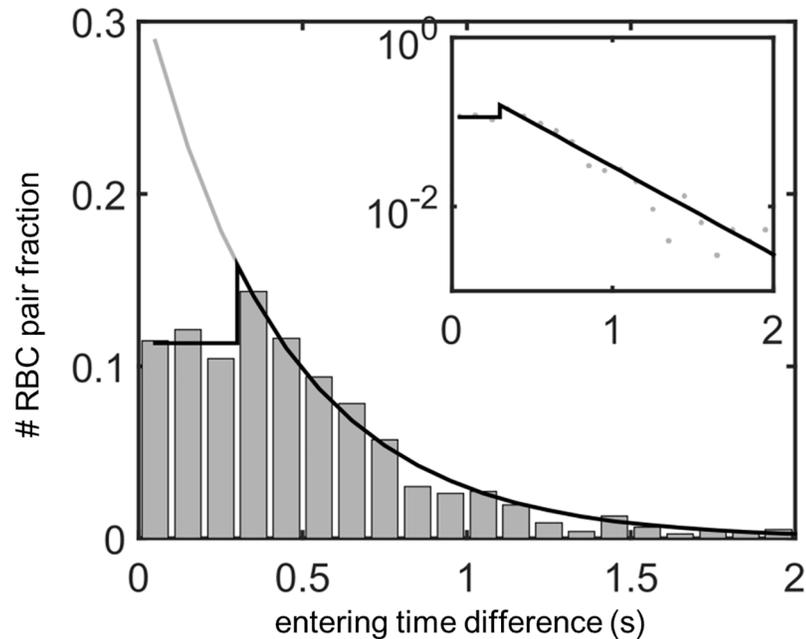

**Fig 3. Red flood cell flows in the real intersegmental artery network are affected by feedbacks, as shown by a significantly lower fraction of red blood cells entering the same vessel within 0.3s of each other.** Shown: Distribution of inter-entry times for cells entering all 12 SeAs. In the absence of feedbacks, inter-entry times will be exponentially distributed (black curve), while real inter-entry times follow an exponential distribution only when cells enter the vessel more than 0.3s apart, and have uniform distribution when cells enter the vessel within 0.3s of each other (black star curve). Inset: The semi-log plot of the linear-exponential distribution (black curve) fits well to the data (gray dots) above 0.3s, showing the exponential distribution when the inter-entry time is long enough for the first cell to leave the vessel. We bin the inter-entry time intervals into 0.1s bins which is the typical time resolution of our videos.

https://doi.org/10.1371/journal.pcbi.1005892.g003

By contrast, we observed large feedback effects within the SeA, i.e. the presence of a red blood cell reduces the flow in the vessel and hence the entering probability of the next cell. We individually tracked red blood cells in a single 4dpf zebrafish, and plotted the inter-entry intervals, i.e. the times between consecutive red blood cells entering each vessel, condensing data from all SeAs since all vessels have the same approximate rate of blood cell entry (see Fig 3). In the absence of feedbacks, we would expect the inter-entry times to be distributed randomly, i.e. as an exponential random variable. Our red blood cell tracking shows that a single red blood cell passes through an SeA in a mean time of 0.3s. Inter-cell entry intervals larger than 0.3s (i.e. cell entries into unoccupied SeAs) were distributed exponentially (see the inset to Fig 3). However, inter-entry intervals less than 0.3s were not exponentially distributed, and we saw far fewer cells entering vessels less than 0.3s apart (i.e. while the vessels were already occupied by other cells) than would be expected based on the exponential distribution (Fig 3, main panel). In fact we found that inter-entry intervals less than 0.3s were approximately uniformly distributed. These observations are suggestive of a negative feedback mechanism, whereby entry of a red blood cell into an SeA reduces for some time afterward the probability of another red blood cell entering the same SeA.

We tested for statistical support for the presence of negative feedback by two methods. First, we extrapolated the exponential fit for time intervals greater than 0.3s to estimate the number of cells that should enter the SeA between 0 and 0.3s, if cell entries into SeA were independent events. For the zebrafish trunk data this amounted to 533 cell entries, compared to the 261 actually observed, and the difference in statistically significant by the Fisher's exact test

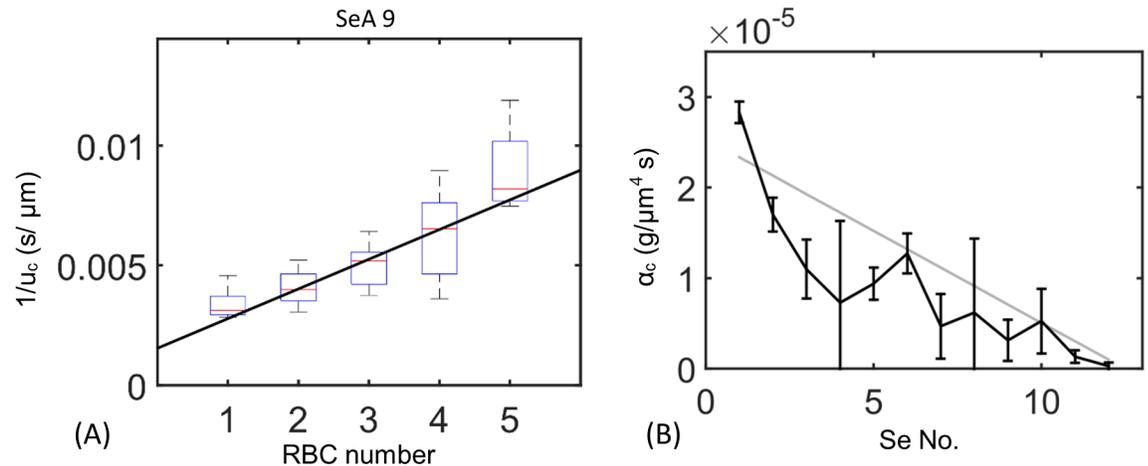

**Fig 4. Occlusion of SeAs by cells feeds back onto the flow through the SeA.** (A) Eq 1 predicts that the reciprocal of cell velocity increases linearly with the number of cells in each Se vessel. Displayed: data from the 9th Se artery (Boxplot) and regression to determine feedback per cell, $\alpha_c$ (curve). The y-intercept is determined from the theoretical plasma velocity in a network with no cells. For data from other Se arteries see S1 Text and S2 Fig. (B) Measured $\alpha_c$ values decrease from first to last Se artery. Gray line: linear regression of $\alpha_c$ against Se vessel index. Bars: 95% confidence intervals calculated by those of linear regressions.

https://doi.org/10.1371/journal.pcbi.1005892.g004

($p = 3.9 \times 10^{-22}$ against independence). Secondly, we fit the distribution of cell entry times directly, to compare an independent model with an exponential probability density function (pdf), with a model in which the feedbacks were modeled by a composite pdf, with uniform probabilities for inter-cell entry intervals less than 0.3s, and an exponential pdf for cell entry intervals greater than 0.3s. The Akaike Information Criterion score corrected for small samples (AICc) [56] for the composite pdf was $4.02 \times 10^3$, whereas the AICc for the pdf assuming independence was $4.07 \times 10^3$, supporting the inclusion of feedback effects.

In mammals red blood cells must squeeze through narrow capillaries. Passage through these narrow vessels is facilitated by specific cellular adaptations—cells are un-nucleated, and have a biconcave shape, assisting cell deformation. By contrast zebrafish red blood cells are almost spherical and are nucleated. However, since the diameters of SeAs are closely comparable to red blood cell diameters (both 6 $\mu$m), we speculated that zebrafish red blood cells may also fit tightly within the SeAs. We directly measured these dynamics by measuring the dependence of the velocity within a SeA upon the number of red blood cells contained in the vessel (see Materials and methods). Velocities within each SeA are affected by the phase of the cardiac cycle as well as by nonlinear cell-cell and cell-wall dynamics [57, 58], so there is large variability in these velocities, and pressures are also affected by changes in conductances throughout the network (Fig 4A). However, in each vessel we found that $1/v$ increased linearly with the number of cells, $n$, consistent with the model for occlusion in Eq (1). In physical terms, when a cell travels through a vessel, it almost blocks the vessel. Because a large pressure difference must be maintained over the cell to push it forward through the SeA, flow within the vessel slows, so that fewer red blood cells enter a vessel once it contains a cell.

We measured the occlusive effect within each SeA, i.e. the parameter $\alpha_c$ in Eq (1) by fitting the slope of the graph of $1/v$ against $n$ (see Fig 4A). The intercept of the graph is given by the speed within the SeA when it contains no red blood cells. We get that speed from the model of flow without occlusive feedbacks, described above, so there is only one free parameter to be estimated for each SeA. Eq (1) represents a form of non-Newtonian rheology, since it gives that the resistance of each vessel increases as hematocrit (i.e. $n$) increases. The parameter $\alpha_c$ represents the intrinsic resistance per cell [34, 35, 37, 38], and it depends on the relative size of the cell and SeA (i.e. how tightly the red blood cell must be squeezed to travel along the vessel), cellular deformation due to elastohydrodynamic effects [32], as well as upon the surface chemistry of both. In particular, [34] built a physically informed model of cells moving through a narrow vessel, including both cell deformation, and interactions between the cell and the vessel glycocalyx: a polymer brush that covers and lubricates the endothelial lining of the vessel. They found that $\alpha_c$ is highly sensitive to biophysical parameters: the thickness of the glycocalyx layer and its porosity (i.e. to the concentration of polymer), as well as to small changes in vessel radius.

To assay the potential for controllability for the occlusive effect, $\alpha_c$, we measured $\alpha_c$ independently in each of the twelve SeAs, in all cases by fitting the data for the dependence of $1/v$ upon $n$ (see S1 Text for more details of the fit). The experimentally measured occlusion strength decreased from first to last SeA (Fig 4B), over a range of $\alpha_c = 3.0 \times 10^{-7} \sim 2.8 \times 10^{-5}$ $g/\mu m^4\ s$. In physical terms, red blood cells occlude closer vessels to the heart more than distal vessels. These values are consistent with the range given in Secomb et al.'s model [34] in which $\alpha_c$ could range from $\alpha_c = 1.8 \times 10^{-7}$ to $1.6 \times 10^{-5}$ $g/\mu m^4\ s$. Our measurement of $\alpha_c$ also agrees with an earlier theoretical model of Secomb et al.'s which did not consider glycocalyx ($\alpha_c = 4.7 \times 10^{-7} \sim 3.8 \times 10^{-6}$ $g/\mu m^4\ s$ [37]), a numerical model of Pozrikidis' which simulated the time course of cell deformation ($\alpha_c = 2.4 \times 10^{-7} \sim 1.1 \times 10^{-6}$ $g/\mu m^4\ s$ [38], as well as an experimental fit to earlier data ($\alpha_c = 1.4 \times 10^{-7}$ $g/\mu m^4\ s$ [36])). Note however, that the micromechanical and numerical models of [34, 37, 38] was created for mammalian red blood cells in capillaries and must be applied with caution here; indeed glycocalyx parameters have not been measured in zebrafish. Although the differences between zebrafish and mammalian RBCs mean that we must allow that the parameters controlling occlusive feedback $\alpha_c$ may be different in zebrafish than in mammalian vessels, the mammalian data generally support the possibility of tuning feedbacks over a large range of values. The intrinsic resistance $\alpha_c$ depends on many factors, including cell velocity, thickness of glycocalyx layer, and the deformation of the cell. Here we focus on the effect of $\alpha_c$ on the partitioning of the cells rather than the detailed mechanism that causes the variation.

## Tuning occlusive effects between different micro-vessels uniformly partitions red blood cells

We simulated around 17 min of red blood cell flow through the zebrafish vascular network, assuming the same occlusive effect for every microvessel, using a discrete model in which every red blood cell trajectory was tracked and in which vessel resistances were modeled using Eq (1) (see Materials and methods) using the same occlusive feedback parameter ($\alpha_c = 1.01 \times 10^{-6}$ $g/\mu m^4\ s$) for each vessel. The model continued to predict that red blood cell fluxes within vessels decrease exponentially with distance from the heart (Fig 1C). This can be rationalized as follows: If $\alpha_c$ is identical between intersegmental vessels, and phase separation is assumed to be negligible, then the model predicts that the resistance of each vessel will increase on average from the value given by the Hagen-Poiseuille law by $\alpha_c \cdot \text{Hct} \cdot V/V_c$, where $V$ is the volume of the vessel, $V_c$ is the volume of a single cell and Hct is the hematocrit. The approximate model derived in Results, Absence of occlusion . . . demonstrates that variation in SeA length from head to tail of the zebrafish contribute very little to partitioning of red blood cell fluxes between SeAs, so changing the resistance of each vessel by an amount simply proportional to its length, will similarly not prevent exponential decay of red blood cell fluxes.

The potential effect size of including occlusive feedbacks is much larger than the effect of phase separation: predicted red blood cell flux decreased by a factor of more than 7 in the

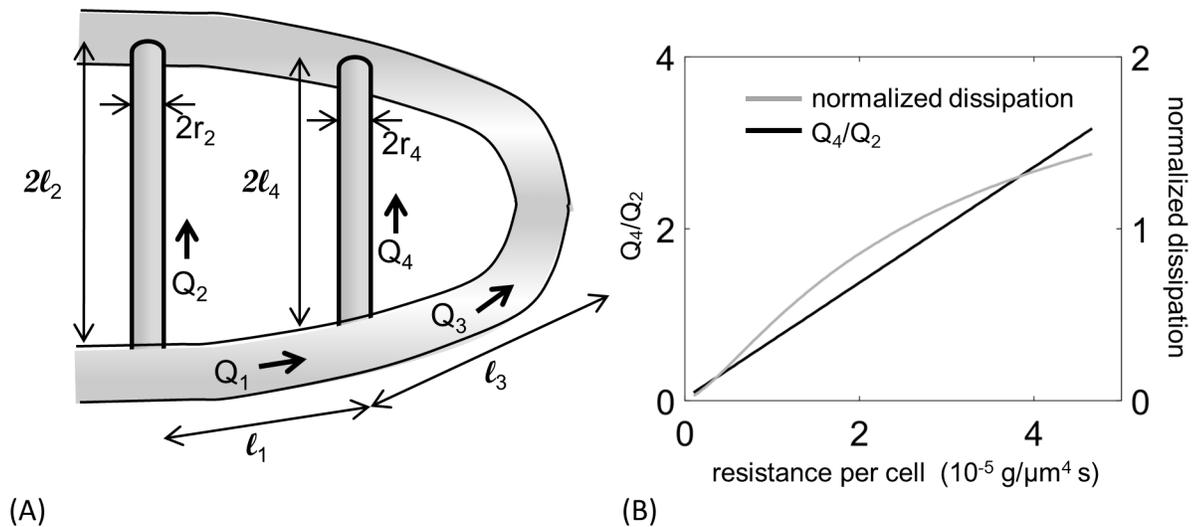

**Fig 5. A reduced vascular network model shows that occlusive effects need to be varied between SeAs, and exposes trade-offs between flow uniformity and transport efficiency.** (A) Diagram of the reduced model of the network showing vessel lengths $l_i$, fluxes $Q_i$, and radii $r_i$. (B) Increasing the occlusion strength $\alpha_2$ increases flux uniformity, measured by the ratio of fluxes in the last and the first Se (black curve), but also increases dissipation (gray curve), if the total flux through both Se vessels is maintained.

https://doi.org/10.1371/journal.pcbi.1005892.g005

phase separation model (see S1 Text). We therefore hypothesized that varying occlusive effects between different SeAs may uniformly distribute red blood cells through the network. To probe how variations in occlusive feedback could be used to control the distribution of red blood cells, we studied a reduced model of the vascular network (readers who are mainly interested in simulation results may skip this analysis by going straight to *Observed variation* in . . .). Specifically, we built a mean field model for the flows in a model network including only the first and last SeAs, as well as the direct connection between the DA and PCVs (the labeling of vessels and branching points is shown in Fig 5A). In each vessel the cells were assumed to be well-mixed and cell fluxes are divided in proportion to flow rates at all nodes. Then the hematocrit will be the same in all vessels. For simplicity we express our equations in terms of the concentration of red blood cells (number / volume), $\rho$, rather than the hematocrit. $\rho$ and hematocrit (Hct) are simply related by $\rho = \text{Hct}/V_c$ where $V_c$ is the volume of a cell. Let $R_i$ be the modified resistance of the $i$th vessel according to Eq (1). Then by applying Kirchoff's first law at the branching points at which first and second Se vessel branch off from the aorta, we obtain the pressures at these points, i.e. $p_1$ and $p_2$:

$$F = \frac{p_1 - p_2}{R_1} + \frac{p_1}{R_2}, \frac{p_1 - p_2}{R_1} = \frac{p_2}{R_3} + \frac{p_2}{R_4}, \tag{4}$$

Here $F$ is the total flux of blood into the network, and we can solve Eq (4) by linear algebra (see S1 Text). Of particular interest is is the ratio of fluxes in the two Se, which measures how uniformly the different vessels are kept supplied with cells:

$$\frac{Q_4}{Q_2} = \frac{R_{2_0} + V_2 \rho \alpha_2}{R_{4_0} + V_4 \rho \alpha_4} \left(1 + \frac{R_1}{R_3} + \frac{R_1}{R_4 + V_4 \rho \alpha_4}\right)^{-1} \tag{5}$$

Here $\alpha_2$, $\alpha_4$ are respectively the values of $\alpha_c$ in the first and last SeA, $R_{2_0}$, $R_{4_0}$ are the resistances of the two SeAs in the absence of red blood cell occlusion, and $V_i$ is the volume of vessel $i$.

Most of the parameters in Eq (5) are tightly constrained: the dimensions of the two Se vessels are similar (in fact $R_{2_0} \approx 2R_{4_0}$ and $V_2 \approx 2V_4$), moreover, since the vessel network extends during development and supplies the tail fin in adult zebrafish [59, 60], the aorta must maintain approximately the same radius along its length, leading to $R_1 \approx 11R_3$. Thus the second factor of Eq (5) $\left(1 + \frac{R_1}{R_3} + \frac{R_1}{R_4 + V_4\rho\alpha_4}\right)^{-1}$ has an upper bound $\frac{1}{12}$. Therefore the only parameters that can be used to increase $Q_4/Q_2$ (i.e. eliminate short-circuiting of the network by the first SeA) are the relative sizes of $\alpha_2$ and $\alpha_4$. $Q_4/Q_2$ is largest if $\alpha_2 \gg \alpha_4$, i.e. if occlusion effects are stronger in the first SeA. Thus uniform flow requires stronger occlusion in vessels close to the heart, consistent with experimental observations in real zebrafish (Fig 4B).

However our reduced model also shows that varying occlusion strengths between vessels creates trade-offs between uniformity and the transport efficiency, measured by the dissipation:

$$D_{\text{network}} = \frac{8\mu_{wb}}{\pi r_a^4}(\ell_1 Q_1^2 + \ell_3 Q_3^2) + \frac{8\mu_{pl}}{\pi r_c^4}(\ell_2 Q_2^2 + \ell_4 Q_4^2) \\ + \rho(Q_2^2 V_2 \alpha_2 + Q_4^2 V_4 \alpha_4). \qquad (6)$$

(See S1 Text for derivation). Here $\ell_i$ is the length of the $i$th vessel, $r_a$ is the radius of the DA, and $r_c$ is the radius of the Se vessels. To compare equivalent networks as we vary $\alpha_2$ we also vary $F$, the total flow into the trunk, to keep the total flux through the pair of Se vessels ($Q_2 + Q_4$) constant. Dissipation in the thin layers of fluid surrounding each RBC dominates $D_{\text{network}}$, so as $\alpha_2$ increases $D_{\text{network}}$ increases. The highest ratios of $Q_4/Q_2$ are therefore also the most dissipative networks (Fig 5B).

## Observed variation in occlusive effects optimizes uniform distribution of red blood cells

We modified our simulation from Results, Tuning occlusive effects... to incorporate the observed variations in occlusive effects; i.e. using the different measured values of $\alpha_c$ in each vessel. We used the regressed data (gray line in Fig 4B) to capture the decreasing trend of $\alpha_c$ from head to tail. When vessels were assigned the experimentally measured values of $\alpha_c$, red blood cells became uniformly distributed between SeAs, and matched closely to the real flow observations (see Fig 2A and 2B).

Are the measured variations in occlusive effects really evidence of adaptive tuning of the zebrafish cardiovascular network, or could they arise from incidental changes caused for example by the different ages of vessels at different distances along the trunk? New SeAs are progressively added to the trunk at the tail of the zebrafish as the trunk elongates, and we wanted to evaluate the alternate hypothesis that the younger vessels farther from the heart had lower occlusive effects simply because they have a thinner glycocalyx coating, or else because structural adaptation of vessels to the flows through them may tend to reduce vessel radii over time [61]. Although neither alternate explanation can be totally ruled out, we were able to test how close the observed distribution of occlusive effects is to one that optimizes the uniform partitioning of red blood cell flows between vessels. Specifically, we ran discrete cell simulations of flow within the network for different distributions of occlusive effects: that is, we varied $\Delta\alpha_c$, defined to be the difference in $\alpha_c$ between the first and last SeAs, assuming a linear variation of $\alpha_c$ in the intermediate vessels. For each model network, we calculated the coefficient of variation (CV) in the red blood cell flux, i.e. the standard deviation in red blood cell flow rate over all vessels, normalized by the mean flow rate. Smaller values of CV correspond to a more uniform distribution of red blood cell flows. Using discrete cell simulations, i.e.

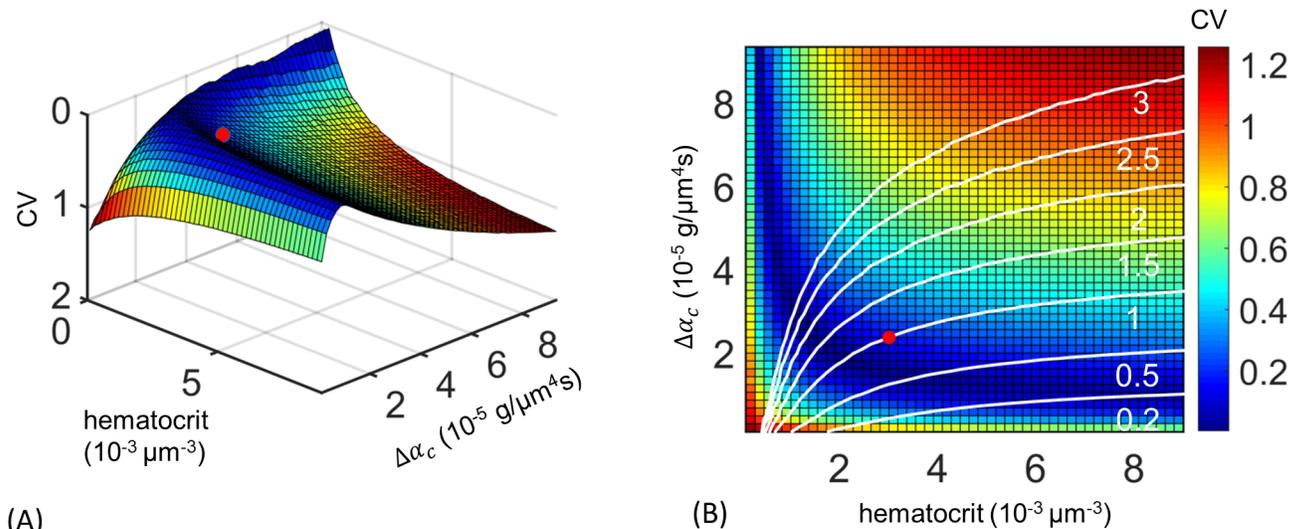

**Fig 6. Tuned occlusion strengths uniformly distribute flow across different Se vessels.** (A) Dependence of flux uniformity upon controllable parameters is explored by allowing blood cell concentration, $\rho$, and difference in occlusive effects between first and last Se vessel, $\Delta\alpha_c$, to vary independently and computing the coefficient of variation (CV) for flow through all Se vessels. Flux uniformity is achieved only within a narrow manifold of values of blood cell concentrations and occlusive effect differences. The empirical values (red dot) lie close to this optimal manifold. (B) Higher uniformity can be achieved if blood cell concentration is decreased (moving leftward from the red dot) but at the cost of increasing dissipation. Transport costs are reduced if $\Delta\alpha_c$ is decreased (moving downward), and can be reduced by 11-fold if there is no difference occlusive effect between different Se arteries but at the cost of reducing uniformity of RBC fluxes. Colors show CV values from (A) and white curves show level sets of dissipation; the dissipation is normalized by its value in the real zebrafish.

https://doi.org/10.1371/journal.pcbi.1005892.g006

tracking every cell trajectory, produces more accurate estimates of red blood cell fluxes in principle than the continuum modeling from Results, Tuning occlusive effects . . ., because cell number fluctuations within each SeA are comparable to the mean number of cells. Since the change in resistance of a vessel depends on the number of cells in the vessel according to Eq (1), the distribution of red blood cell flows for a given distribution of occlusive effects depends on hematocrit. Accordingly, we varied both hematocrit and occlusive effect distributions independently in our simulations. We found for any fixed hematocrit, near uniform flux (CV close to 0) can be achieved only over a narrow range of $\Delta\alpha_c$ (Fig 6A). Too little difference in intrinsic resistance between first and last SeAs, and the first SeA short-circuits the network, as discussed in Results, Absence of occlusion . . .. But too large a difference in occlusive effects can have the opposite effect, leading to the vessels furthest from the heart receiving more flow than vessels closest to the heart. The optimal distribution the occlusive effects is realized along a single curve in ($\Delta\alpha_c$, $\rho$) space. We found that the observed occlusion effect distribution is close to the optimal value for the real zebrafish hematocrit [43] (Fig 6A).

## Discussion

Our work shows that feedbacks associated with the occlusion of fine vessels by the red blood cells that pass through may be associated with previously unreported adaptive benefits for control of blood flows within the microvasculature. Although the existence of occlusive feedbacks is well known [54, 58, 62, 63], to our knowledge they have not previously been shown to be associated with adaptive benefits for oxygen perfusion. Although our experimental observations and modeling are focused on zebrafish, which are a model for vascular development, it is likely that similar feedbacks are significant within mammalian microcirculatory systems, where the deformation of cells to pass through capillaries is, if anything, even more extreme than in the zebrafish. Indeed the apparent intrinsic resistance of cells in human blood vessels has a wide range of variability [34, 37], and precise tuning of blood flows is already known to be vital e.g. to maintain perfusion-ventilation balance in the lungs [64–66]. The proposed occlusion feedback mechanism may be able to explain the variation of capillary blood flow and how it affects the ventilation-perfusion ratio, as well as blood flows in other vascular systems such as brain capillary network.

Capillary networks have been hypothesized to be organized to minimize the cost of blood transport [12, 13]. Although large vessels seem to conform very closely to this organizing principle [13, 24], the tuning of occlusive effects to uniformly distribute red blood cell flows takes the zebrafish vascular network far from the configuration that minimizes transport costs. In particular, at the physiological hematocrit, if the same (smallest) occlusive effect, $\alpha_c$, is assigned to each vessel then the dissipation in the network could be reduced by a factor of 11 (Fig 6B). At the same time, more uniform partitioning of cell fluxes between different SeAs (i.e. a lower value of the Coefficient of Variation of red blood cell flow rates) is possible but altering physiological parameters further decreases the transport efficiency. For example decreasing blood cell concentration, $\rho$, increases uniformity of flux, but at the cost of increasing dissipation if the total cell supply to all Se vessels is to kept fixed (Fig 6B).

The ability of SeAs to vary the occlusive effect $\alpha_c$ over three orders of magnitude is consistent with previous modeling of red blood cell and microvessel mechanics, and endows the network with tremendous control over red blood cell flow rates. It is natural to ask whether and how uniform red blood cell flux partitioning can be maintained against the numerous sources of perturbation that occur in real cardiovascular networks. Microvascular networks may be disrupted by trauma, micro-anneurysms, or by systemic conditions like diabetes mellitus [67–70]. As a first step toward answering this question, we considered the effect of well-characterized natural variability in SeA spacing [48], and of the *notch* mutation which alters the trunk network connectivity [71] upon the ability of the network to uniformly distribute red blood cell fluxes. We found that under a wide range of vessel spacing variability, red blood cell fluxes remained uniform across all SeAs (see S1 Text and S3 Fig). Indeed vessel spacing variability has no detectable effect on zebrafish growth and maturation. By contrast, in notch mutant zebrafish the cardiovascular network is malformed, with a shunt connection forming between aorta and principal cardinal vein (S4 Fig). Since the diameter of the shunt is much larger than the cell diameter, there is negligible occlusive feedback within the shunt, causing it to irreparably short-circuit the vascular network. Shunt formation is lethal in embryos, and our model shows that it creates conditions under which uniform perfusion of the trunk is impossible. Note however that mechanisms not described in the model can still play significant roles in both developmental process and mutant network phenotypes. For example in the *gridlock* mutant [51] the blood flow to the tail is impeded by a localized vascular defect, but the collateral vessels not present at 4dpf was observed to redirect the flow around the blockade and rescue the embryo. During the development process both the number of vessels and size of zebrafish embryo change dramatically. Therefore we expect an observable change in occlusive feedbacks to maintain uniform cell partition throughout the developmental stages. Extending our analysis to include the topological changes observed as embryonic zebrafish develop [51] is an ongoing effort.

Although we are able to directly demonstrate that occlusive feedbacks vary between different the SeAs, and this variation is consistent with optimization of feedback strengths to ensure uniform distribution of red blood cells across trunk vessels, our model cannot reveal what physical changes within vessels are used with the zebrafish network to modulate the occlusive effect. In our experiments we cannot visualize the glycocalyx lining of the SeAs, and in fact we are aware of no previous works in which glycocalyx was measured in blood vessels simultaneously with flow. However, previous studies have reported large variations in glycocalyx porosity and thickness between different vessels [32, 72]. Since cells must squeeze into SeAs, variations in vessel radius below the resolution limit of our microscopy method could also account for the variation in occlusive effect. Finally elastohydrodynamic effects associated e.g. with changes in the speed of cells, [32], may affect feedback models. The analysis is also silent on the mechanisms for coordinating occlusive effects across the network. Recent works have dissected structural adaptations in microvascular networks [61], as well as in biological transport networks generally [73–75]. These works have focused on the question of how a set of vascular elements that have information only about their own flows can alter their resistances in response to these cues to minimize dissipation within the network. This question is directly relevant to other objective functions i.e. to networks that maximize uniformity rather than maximizing hydraulic efficiency—can vessels adapt their occlusive effects to the their flow to achieve uniform red blood cell transport?

The use of tuned occlusive effects creates uniform distribution of red blood cell fluxes through the zebrafish vascular network, but at the cost of increasing transport costs. Indeed if the network simply used the same value of $\alpha_c$ in every SeA we found that an 11 fold decrease in transport costs would be possible within the zebrafish trunk vasculature (Fig 6B). Physically feedbacks from occlusion represent a form of congestion, and efficient transport networks, both natural [76] and artificial [10, 11], are often organized to avoid congestion. Previous works have provided algorithms for constructing minimally dissipative networks given a prescribed set of sources and sinks [22, 23]. Our work suggests that other optimizing principles may govern microvascular network organization. Extending network optimization algorithms to include flow uniformity is likely to further reveal the tradeoffs between uniformity and efficiency.

## Supporting information

**S1 Fig. Phase separation model.**
(TIF)

**S2 Fig. Regression on red blood cell velocity data for all 12 Se vessels in a 4dpf zebrafish.**
(TIF)

**S3 Fig. Predicted cell fluxes with variability in Se spacing.**
(TIF)

**S4 Fig. Predicted cell flux in *notch* mutant zebrafish.**
(TIF)

**S1 Table. Data on zebrafish geometry.**
(PDF)

**S1 Text. Oxygenation model, phase separation model, derivation of the mean field model, and the effect of network perturbation on red blood cell partitioning.**
(PDF)

## Acknowledgments

We thank the Taiwan Zebrafish Core Facilities at Academia Sinica (TZCAS) and at NHRI (TZCF) for providing Tg(gata1:dsRed; fli1a:EGFP) transgenic fish, and Jérémy Lévéque for performing early calculations for Fig 1C.

## Author Contributions

**Conceptualization:** Shyr-Shea Chang, Shenyinying Tu, Van M. Savage, Marcus Roper.

**Data curation:** Shyr-Shea Chang, Shenyinying Tu.

**Formal analysis:** Shyr-Shea Chang, Marcus Roper.

**Funding acquisition:** Sheng-Ping L. Hwang, Tzung K. Hsiai, Marcus Roper.

**Investigation:** Kyung In Baek, Andrew Pietersen, Yu-Hsiu Liu.

**Methodology:** Shyr-Shea Chang, Shenyinying Tu, Marcus Roper.

**Project administration:** Sheng-Ping L. Hwang, Tzung K. Hsiai, Marcus Roper.

**Resources:** Kyung In Baek, Andrew Pietersen, Yu-Hsiu Liu, Sheng-Ping L. Hwang, Tzung K. Hsiai, Marcus Roper.

**Software:** Shyr-Shea Chang, Shenyinying Tu, Marcus Roper.

**Supervision:** Sheng-Ping L. Hwang, Tzung K. Hsiai, Marcus Roper.

**Validation:** Shyr-Shea Chang, Shenyinying Tu.

**Visualization:** Shyr-Shea Chang, Shenyinying Tu, Marcus Roper.

**Writing – original draft:** Shyr-Shea Chang, Shenyinying Tu, Marcus Roper.

**Writing – review & editing:** Shyr-Shea Chang, Marcus Roper.

PLOS COMPUTATIONAL BIOLOGY                                                Optimal occlusion and uniform partition of RBC13. Sherman TF. On connecting large vessels to small. The meaning of Murray's law. The Journal of general physiology. 1981; 78(4):431–453. https://doi.org/10.1085/jgp.78.4.431 PMID: 7288393

14. Rao S, Bálint Š, Cossins B, Guallar V, Petrov D. Raman study of mechanically induced oxygenation state transition of red blood cells using optical tweezers. Biophysical journal. 2009; 96(1):209–216. https://doi.org/10.1529/biophysj.108.139097 PMID: 18931252

15. Wang CH, Popel AS. Effect of red blood cell shape on oxygen transport in capillaries. Mathematical biosciences. 1993; 116(1):89–110. https://doi.org/10.1016/0025-5564(93)90062-F PMID: 8343620

16. Kleinfeld D, Mitra PP, Helmchen F, Denk W. Fluctuations and stimulus-induced changes in blood flow observed in individual capillaries in layers 2 through 4 of rat neocortex. P Natl Acad Sci USA. 1998; 95 (26):15741–15746. https://doi.org/10.1073/pnas.95.26.15741

17. Chaigneau E, Oheim M, Audinat E, Charpak S. Two-photon imaging of capillary blood flow in olfactory bulb glomeruli. P Natl Acad Sci USA. 2003; 100(22):13081–13086. https://doi.org/10.1073/pnas.2133652100

18. Schwerte T, Überbacher D, Pelster B. Non-invasive imaging of blood cell concentration and blood distribution in zebrafish Danio rerio incubated in hypoxic conditions in vivo. J Exp Biol. 2003; 206(8):1299–1307. https://doi.org/10.1242/jeb.00249 PMID: 12624165

19. Pries A, Secomb TW, Gaehtgens P. Structure and hemodynamics of microvascular networks: heterogeneity and correlations. American Journal of Physiology-Heart and Circulatory Physiology. 1995; 269 (5):H1713–H1722.

20. Chico TJ, Ingham PW, Crossman DC. Modeling cardiovascular disease in the zebrafish. Trends in cardiovascular medicine. 2008; 18(4):150–155. https://doi.org/10.1016/j.tcm.2008.04.002 PMID: 18555188

21. Banavar JR, Colaiori F, Flammini A, Maritan A, Rinaldo A. Topology of the fittest transportation network. Physical Review Letters. 2000; 84(20):4745. https://doi.org/10.1103/PhysRevLett.84.4745 PMID: 10990786

22. Bohn S, Magnasco MO. Structure, scaling, and phase transition in the optimal transport network. Physical review letters. 2007; 98(8):088702. https://doi.org/10.1103/PhysRevLett.98.088702 PMID: 17359138

23. Katifori E, Szöllősi GJ, Magnasco MO. Damage and fluctuations induce loops in optimal transport networks. Phys Rev Lett. 2010; 104(4):048704. https://doi.org/10.1103/PhysRevLett.104.048704 PMID: 20366746

24. Zamir M, Sinclair P, Wonnacott TH. Relation between diameter and flow in major branches of the arch of the aorta. Journal of biomechanics. 1992; 25(11):1303–1310. https://doi.org/10.1016/0021-9290(92)90285-9 PMID: 1400531

25. Schindelin J, Arganda-Carreras I, Frise E, Kaynig V, Longair M, Pietzsch T, et al. Fiji: an open-source platform for biological-image analysis. Nature methods. 2012; 9(7):676–682. https://doi.org/10.1038/nmeth.2019 PMID: 22743772

26. Hove JR, Köster RW, Forouhar AS, Acevedo-Bolton G, Fraser SE, Gharib M. Intracardiac fluid forces are an essential epigenetic factor for embryonic cardiogenesis. Nature. 2003; 421(6919):172–177. https://doi.org/10.1038/nature01282 PMID: 12520305

27. Jones EA, le Noble F, Eichmann A. What determines blood vessel structure? Genetic prespecification vs. hemodynamics. Physiology. 2006; 21(6):388–395. https://doi.org/10.1152/physiol.00020.2006 PMID: 17119151

28. Pedley TJ. Mathematical modelling of arterial fluid dynamics. Journal of engineering mathematics. 2003; 47(3-4):419–444. https://doi.org/10.1023/B:ENGI.0000007978.33352.59

29. Windberger U, Bartholovitsch A, Plasenzotti R, Korak K, Heinze G. Whole blood viscosity, plasma viscosity and erythrocyte aggregation in nine mammalian species: reference values and comparison of data. Exp Physiol. 2003; 88(3):431–440. https://doi.org/10.1113/eph8802496 PMID: 12719768

30. Lighthill J. Mathematical biofluiddynamics. vol. 17. Siam; 1975.

31. Malone MH, Sciaky N, Stalheim L, Hahn KM, Linney E, Johnson GL. Laser-scanning velocimetry: a confocal microscopy method for quantitative measurement of cardiovascular performance in zebrafish embryos and larvae. BMC biotechnology. 2007; 7(1):40. https://doi.org/10.1186/1472-6750-7-40 PMID: 17623073

32. Weinbaum S, Tarbell JM, Damiano ER. The structure and function of the endothelial glycocalyx layer. Annu Rev Biomed Eng. 2007; 9:121–167. https://doi.org/10.1146/annurev.bioeng.9.060906.151959 PMID: 17373886

33. Pries AR, Secomb TW. Microvascular blood viscosity in vivo and the endothelial surface layer. American Journal of Physiology-Heart and Circulatory Physiology. 2005; 289(6):H2657–H2664. https://doi.org/10.1152/ajpheart.00297.2005 PMID: 16040719
PLOS Computational Biology | https://doi.org/10.1371/journal.pcbi.1005892   December 15, 2017   20 / 22